\documentclass{jfm}

\usepackage{graphicx,subfigure}
\usepackage{amsbsy,amssymb,amsmath,bm}
\usepackage{natbib}

\title{Stratified precessional flow in spherical geometry}
\author[Xing Wei and Andreas Tilgner]{XING WEI\thanks{Email address for correspondence: xing.wei@phys.uni-goettingen.de}, ANDREAS TILGNER}
\affiliation{Institute of Geophysics, University of G\"ottingen\\Friedrich-Hund-Platz 1, G\"ottingen 37077, Germany}
\pubyear{?}
\volume{?}
\pagerange{?--?}
\date{?; revised ?; accepted ?.}

\begin{document}
\maketitle

\begin{abstract}
We investigate numerically in spherical geometry the interaction of stratification with precession. Both stable stratification and unstable stratification are studied. In the parameter regime we are concerned with, stable stratification suppresses the precessional instability, whereas unstable stratification and precession can either stablise or destablise each other at the different precession rates.
\end{abstract}


\section{Introduction and motivation}
The magnetic field of many celestial bodies is maintained by the dynamo effect which requires that the body contains a fluid conductor executing a non-trivial motion. The energy source for this motion is frequently assumed to be thermal convection. Accordingly, convection driven dynamos in spherical spheres have been extensively studied \citep{busse1,roberts1,wicht,jones}. \citet{bullard} first pointed out that precession is a possible energy source for the geodynamo. Precession driven flows, their stability and dynamo properties have been investigated independently of thermal convection \citep{poincare,malkus,busse2,vanyo,tilgner4,noir,tilgner1,tilgner7,tilgner3,tilgner2}. More recently, some attention has also been paid to other types of mechanical forcing, such as tides, libration and collisions \citep{weiss,bars1,dwyer}.

For the Earth and the other planets and moons of the solar system, the mechanical forcing is well known, while it is uncertain if their cores are convecting. In general, it will always be necessary to consider buoyancy and a mechanical forcing such as precession in conjunction. The thermal stratification can be either unstable or stable depending on which epoch in the thermal history of a celestial body is being considered. Typically, the surface cooling of a recently formed body leads to a superadiabatic gradient. As the heat generated during accretion and formation of the body is progressively lost, the internal temperature gradient eventually drops below the adiabatic gradient, yielding a stable thermal stratification. Small bodies such as planetesimals or the Moon have already reached this point in the past.

In this paper, we investigate the hydrodynamics of stably and unstably stratified fluid in a precessing spherical shell. This is intended as a model of the Earth, whose precession is maintained by the gravitational torque of the Moon and the Sun, but also as a model of planetesimals in the early solar system which undergo force free precessional motion after collisions if the direction of their angular momentum does not coincide with one of their principal axes of inertia.

In section 2 we formulate the problem and introduce the numerical methods. In section 3 we investigate the interaction of stable stratification and precession. In section 4 we investigate the interaction of unstable stratification and precession. We conclude with some discussion in section 5.

\section{Equations and numerical methods}
We consider a fluid in a spherical shell of inner radius $r_i$ and outer radius $r_o$. Suppose that the spherical shell spins at the angular velocity $\bm\Omega_s$ and precesses at the angular velocity $\bm\Omega_p$. A background temperature $T_b$ is imposed on the fluid, which we assume to be a linear function of radius,
\begin{equation}\label{tb}
T_b=\frac{T_o-T_i}{d}(r-r_o)+T_o,
\end{equation}
where $T_i$ and $T_o$ are the temperature at $r_i$ and $r_o$, respectively, and $d=r_o-r_i$ is the thickness of the gap. Such a background temperature is maintained by a heat source or sink which varies inversely proportional to radius (because $\nabla^2 T_b = 2(T_o-T_i)/(rd)$). An alternative model in which the stratification is imposed by fixed temperatures at the boundaries in the absence of a heat source leads to a non-uniform stratification and has not been considered for simplicity. In the frame moving with the boundary, the dimensional Navier-Stokes equation of a Boussinesq fluid reads
\begin{align}\label{ns1}
&\frac{\partial\bm u}{\partial t}+\bm u\bm\cdot\bm\nabla\bm u+\bm\Omega_{ref}\times(\bm\Omega_{ref}\times\bm r)+2\bm\Omega_{ref}\times\bm u+(\bm\Omega_p\times\bm\Omega_s)\times\bm r= \nonumber\\
&-\frac{1}{\rho_o}\bm\nabla p+\nu\nabla^2\bm u+\frac{\rho}{\rho_o}\bm g, ~~~ \nabla \cdot \bm u =0,
\end{align}
where $\bm\Omega_{ref}=\bm\Omega_s+\bm\Omega_p$ is the angular velocity of the moving frame with respect to the inertial frame and $\rho_o$ the fluid density at $r_o$. In the Boussinesq approximation, density variations are only retained in the buoyancy term and the density deviation is proportional to the temperature deviation,
\begin{equation}\label{boussinesq}
\frac{\rho-\rho_o}{\rho_o}=-\alpha(T-T_o)=-\alpha(\Theta+T_b-T_o),
\end{equation}
where $\Theta=T-T_b$ is the temperature deviation from the conductive profile and $\alpha$ the thermal expansion. In a sphere of constant density, the gravitational acceleration is proportional to radius,
\begin{equation}\label{g}
\bm g=-g_o\frac{\bm r}{r_o},
\end{equation}
where $g_o$ is the gravity at $r_o$. 

Substituting (\ref{boussinesq}) and (\ref{g}) into (\ref{ns1}) and normalising length with $d$, time with $\Omega_s^{-1}$, $\bm u$ with $\Omega_sd$ and $\Theta$ with $(T_i-T_o)$, we obtain the dimensionless Navier-Stokes equation
\begin{equation}\label{ns2}
\frac{\partial\bm u}{\partial t}+\bm u\bm\cdot\bm\nabla\bm u=-\bm\nabla\Phi+Ek\nabla^2\bm u+2\bm u\times(\hat{\bm z}+Po\hat{\bm\Omega}_p)+Po(\hat{\bm z} \times\hat{\bm\Omega}_p)\times\bm r+\widetilde{Ra}\Theta\bm r,
\end{equation}
where all the curl-free terms, e.g. the centrifugal force and the term associated with $T_b$, are already absorbed into the total pressure $\Phi$. In (\ref{ns2}) there are three dimensionless parameters. The Ekman number 
\begin{equation}\label{ek}
Ek=\frac{\nu}{\Omega_sd^2}
\end{equation}
measures the ratio of the viscous time scale to the spin time scale, the Poincar\'e number
\begin{equation}\label{po}
Po=\frac{\Omega_p}{\Omega_s}
\end{equation}
measures the precession rate. Despite its unusual form, we choose to call the parameter in the buoyancy term the Rayleigh number
\begin{equation}\label{ra}
\widetilde{Ra}=\frac{\alpha g_o(T_i-T_o)}{\Omega_s^2r_o},
\end{equation}
which is proportional to the imposed stratification. A negative $\widetilde{Ra}$ corresponds to a stable stratification and a positive $\widetilde{Ra}$ to an unstable stratification. For a stable stratification $\widetilde{Ra}$ is just the square of the dimensionless Brunt-V\"ais\"al\"a frequency. The unit vector along the precession axis in Cartesian coordinates $(x,y,z)$ is
\begin{equation}\label{omega_p}
\hat{\bm\Omega}_p=\sin\beta\cos t\,\hat{\bm x}-\sin\beta\sin t\,\hat{\bm y}+\cos\beta\,\hat{\bm z},
\end{equation}
where $\hat{\bm z}$ is the spin axis and $\beta$ the angle between the spin axis $\hat{\bm z}$ and the precession axis $\hat{\bm\Omega}_p$. The boundary condition of velocity is no-slip at the outer boundary, whereas it is stress-free at the inner boundary to approximate a full sphere.

Accordingly, the dimensionless equation of temperature deviation is
\begin{equation}\label{theta}
\frac{\partial\Theta}{\partial t}+\bm u\bm\cdot\bm\nabla\Theta-u_r=\frac{Ek}{Pr}\nabla^2\Theta.
\end{equation}
In (\ref{theta}) the Prandtl number
\begin{equation}\label{pr}
Pr=\frac{\nu}{\kappa}
\end{equation}
measures the ratio of viscosity to the thermal diffusivity. The boundary condition is $\Theta=0$ at both inner and outer boundaries.

Equations (\ref{ns2}) and (\ref{theta}) are solved numerically. In the calculations we fix the aspect ratio $r_i/r_o$ to $0.1$ such that the spherical shell is almost a spherical cavity and hence the inner core is almost negligible, and the angle $\beta$ to $60^\circ$. We investigate only retrograde precession ($Po<0$) which is geophysically relevant. We fix the Prandtl number to $Pr=1$. In the calculations of stable stratification we fix $Po=-0.3$ and vary $Ek$ and $\widetilde{Ra}$ to investigate how the stable stratification influences the precessional instability. In the calculations of unstable stratification, we select two Ekman numbers $Ek=5\times10^{-3}$ and $1\times10^{-3}$, which are sufficiently high to maintain the flow precessionally stable, and vary $Po$ and $\widetilde{Ra}$ to investigate the interaction of precessional and convective instabilities.

The numerical calculations are carried out in the spherical coordinates $(r,\theta,\phi)$ with the parallel pseudo-spectral code provided by \citet{tilgner5}. The toroidal-poloidal decomposition method is employed such that the divergence free condition of fluid flow $\bm\nabla\bm\cdot\bm u=0$ is automatically satisfied. The spherical harmonics $P_l^m(\cos\theta)e^{{\mathrm i}m\phi}$ are used on the spherical surface $(\theta,\phi)$ and the Chebyshev polynomials $T_k(r)$ are used in the radial direction. Resolutions as high as $256$ (radial $r$), $128$ (colatitude $\theta$) and $64$ (longitude $\phi$) are used. A semi-implicit scheme is employed for time stepping, using an Adams-Bashforth scheme for the nonlinear and a Crank-Nicolson scheme for the diffusive terms.

\section{Stably stratified precessional flow}
The precessional flow in the interior of spherical container is mainly a solid body rotation with angular velocity $\bm\Omega_f$ which is in general different from both $\bm\Omega_s$ and $\bm\Omega_p$ \citep{busse2}. At the boundary the fluid rotation matches the spin rate $\bm\Omega_s$ such that there exists an  Ekman boundary layer, where the poloidal flow is generated by Ekman pumping. In addition to the boundary layers there also exist internal shear layers that are spawned at the critical latitude \citep{tilgner4}. A well-understood instability mechanism in these flows is a triad resonance between the basic flow and two inertial modes \citep{kerswell1,tilgner1,tilgner7,tilgner3}.

We now consider the interaction of a stable stratification and precession. Radial stratification tends to suppress fluid motion in the radial direction, which suggests that stable stratification suppresses precessional instability. However, stable stratification leads to gravito-inertial waves whose frequencies may be closer to a perfect triad resonance than the pure inertial waves of the unstratified case, so that destabilisation through stable stratification is possible, too \citep{kerswell2}. Only stabilisation is observed in our numerical calculations. In the numerical calculations we fix $Po=-0.3$ and calculate various combinations of $Ek$ and $\widetilde{Ra}$, namely $Ek$ ranges from $5\times10^{-3}$ to $1\times10^{-4}$ and $\widetilde{Ra}$ from $0$ (purely precessional flow) to $-1$. Because a stable precessional flow is anti-symmetric about the centre, i.e. $\bm u(-\bm r)=-\bm u(\bm r)$, we use symmetries \citep{tilgner7} to detect the precessional instability, i.e. the flow is separated into two parts, $\bm u_s(\bm r)=\left[\bm u(\bm r)-\bm u(-\bm r)\right]/2$ and $\bm u_a(\bm r)=\left[\bm u(\bm r)+\bm u(-\bm r)\right]/2$, and the kinetic energy of $\bm u_a$ represents the instability.

\begin{figure}
\begin{center}
\subfigure[]
{\includegraphics[scale=0.46]{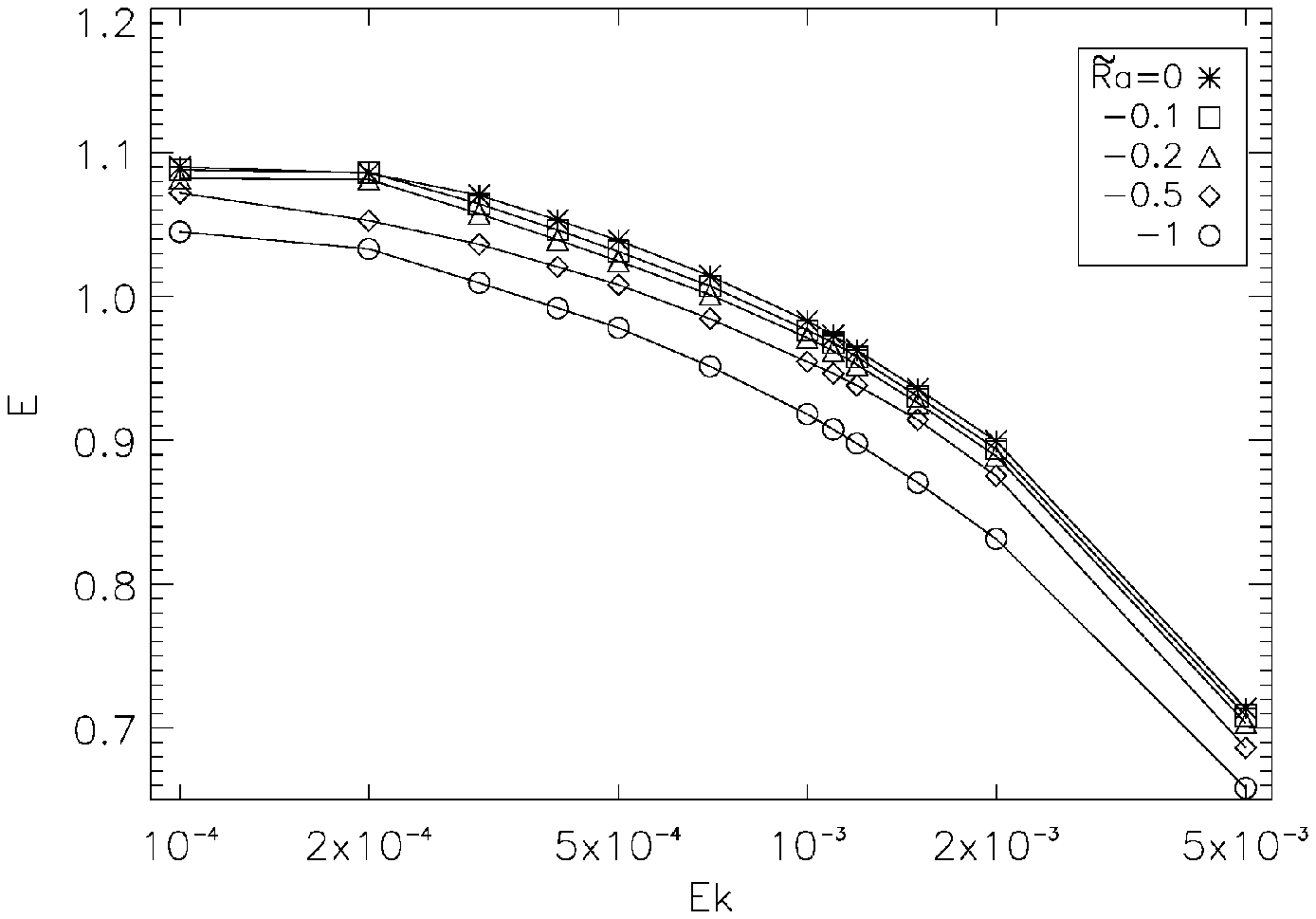}\label{E}}
\subfigure[]
{\includegraphics[scale=0.46]{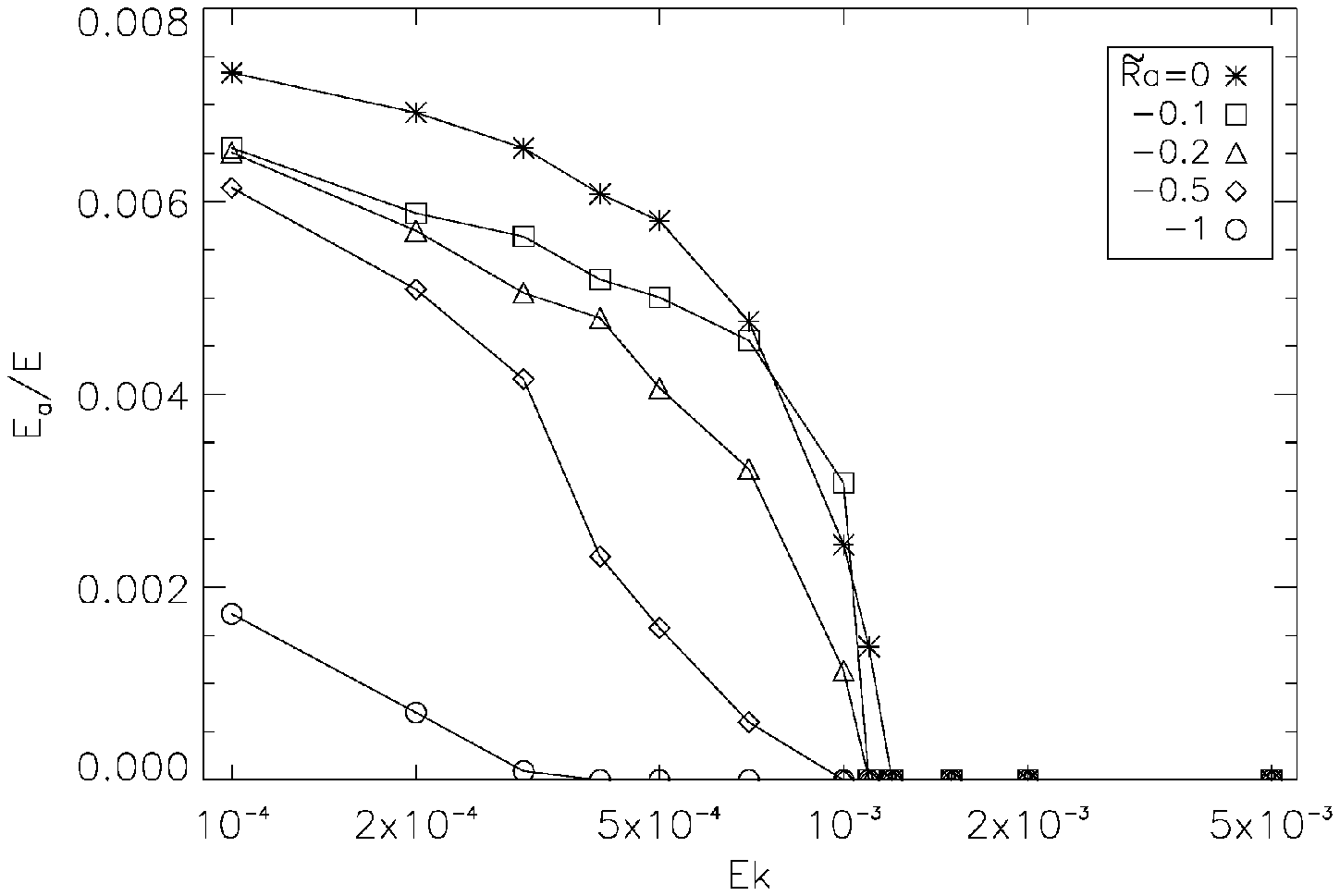}\label{Ea}}
\end{center}
\caption{\footnotesize Stably stratified precessional flow at Poincar\'e number $Po=-0.3$. The total kinetic energy $E$ (a) and the ratio of instability energy $E_a$ to total energy (b) as a function of Ekman number $Ek$ at different Rayleigh numbers $\widetilde{Ra}$.}\label{energy}
\end{figure}

\begin{figure}
\begin{center}
\subfigure[]
{\includegraphics[scale=0.46]{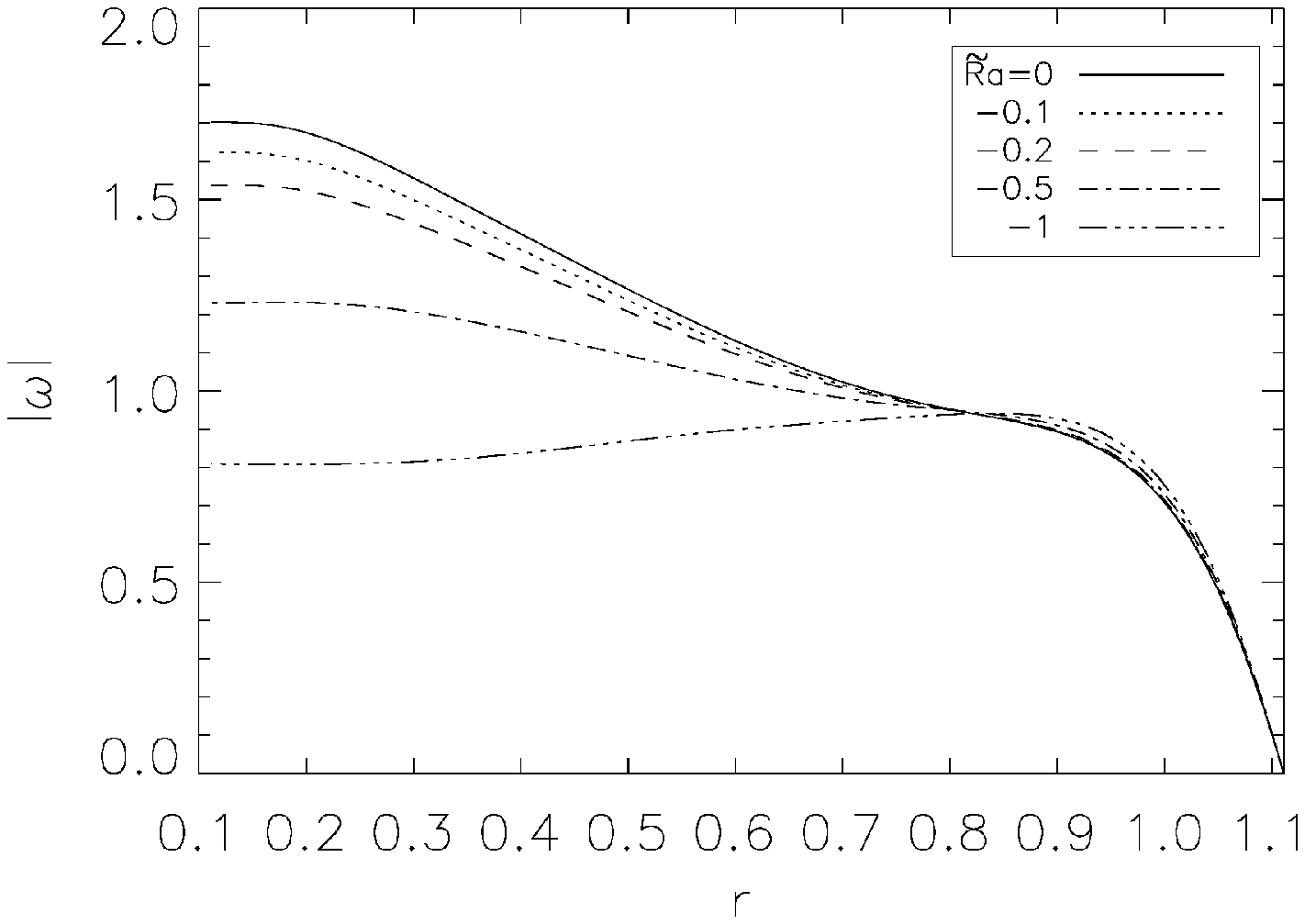}\label{omega1}}
\subfigure[]
{\includegraphics[scale=0.46]{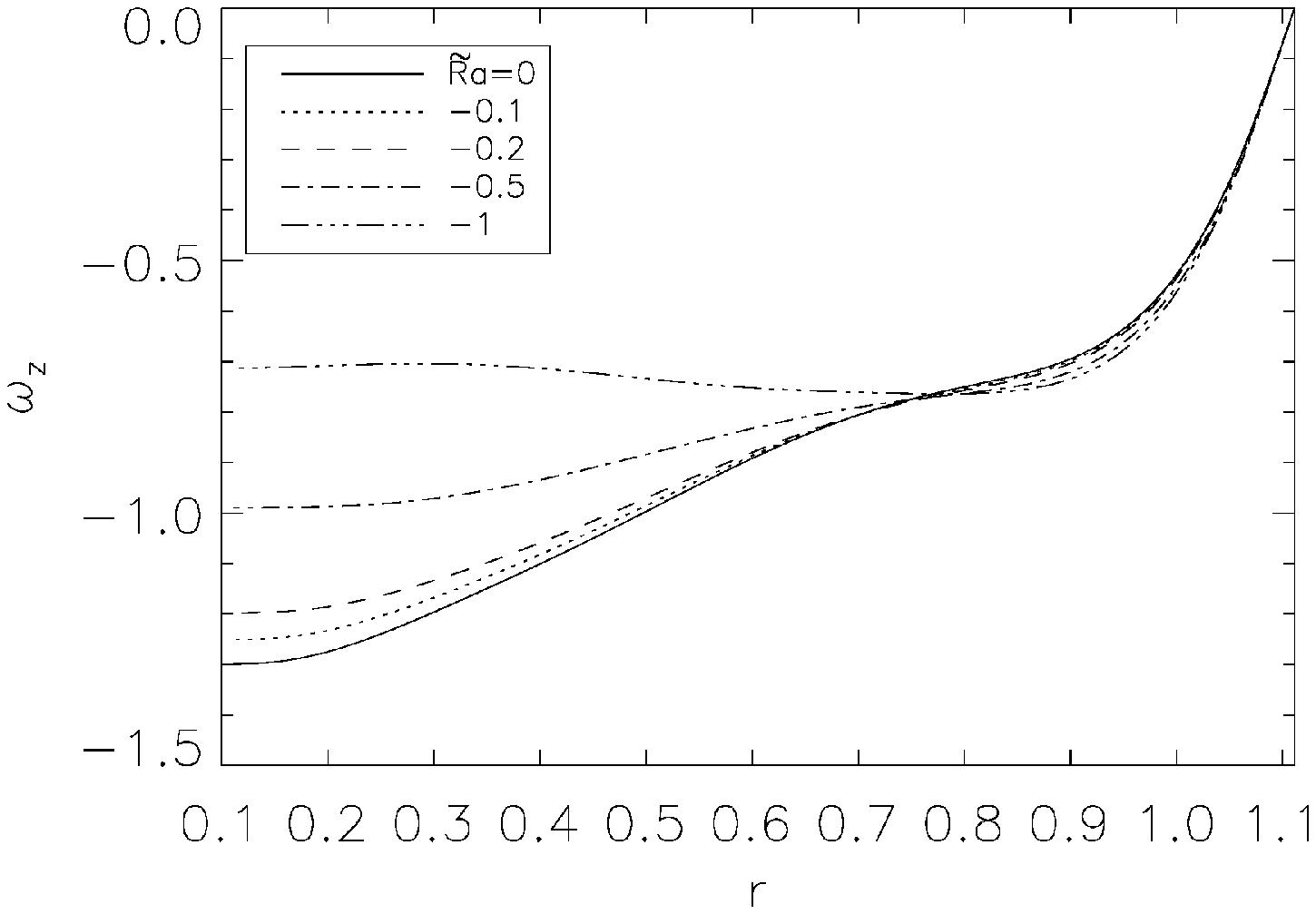}\label{omega2}}
\end{center}
\caption{\footnotesize Stably stratified precessional flow. Radial dependence of fluid rotation vector at Ekman number $Ek=2\times10^{-3}$ and Poincar\'e number $Po=-0.3$. Modulus $|\bm\omega|$ (a) and $z$ component $\omega_z$ (b).}\label{omega}
\end{figure}

Figure \ref{energy} shows the total kinetic energy $E$ (figure \ref{E}) and the ratio of instability energy $E_a$ to the total energy (figure \ref{Ea}) against the Ekman number $Ek$ at different Rayleigh numbers $\widetilde{Ra}$ (for time dependent flows, $E$ and $E_a$ are calculated by averaging over time). Figure \ref{E} shows that increasing $Ek$ reduces the energy $E$. This behavior is well known from previous studies and is due to the fact that the viscous term is more important at higher $Ek$ which damps any motion excited by the Poincar\'e force. It is also seen from figure \ref{E} that a larger $\widetilde{|Ra|}$ reduces the energy. This is consistent with figure \ref{omega1}, which shows the modulus of the fluid rotation vector $|\bm\omega|=\sqrt{\omega_x^2+\omega_y^2+\omega_z^2}$ to indicate that a lower $\widetilde{|Ra|}$ corresponds to a stronger fluid rotation. In order to explain the data in figures \ref{E} and \ref{omega}, one would have to extend Busse's calculation \citep{busse2} to a stably stratified medium, which is not attempted here.

Figure \ref{omega} shows the radial dependence of the fluid rotation vector. The shear in the basic flow is due to viscous corrections to the solution of the inviscid equation, which is simply a solid-body rotation. Viscosity introduces Ekman pumps and shear layers crossing the entire fluid volume, all of which possess a radial velocity component. Since stratification suppresses radial motion, it also reduces the shear. In addition to the modulus of the fluid rotation vector, we also calculated the angle $\gamma$ between the fluid rotation vector and its average, $\gamma=\arccos(\bm\omega\cdot\overline{\bm\omega}/(|\bm\omega||\overline{\bm\omega}|)$, and $\gamma$ is less than around $10^\circ$ in the interior and reaches around $30^\circ$ near the boundary. Moreover, it is interesting that figure \ref{omega2} indicates a counter-rotation of the fluid in the precession frame ($\omega_z+1<0$), which is located in the vicinity of the inner boundary.

\begin{figure}
\begin{center}
\subfigure[]
{\includegraphics[scale=0.46]{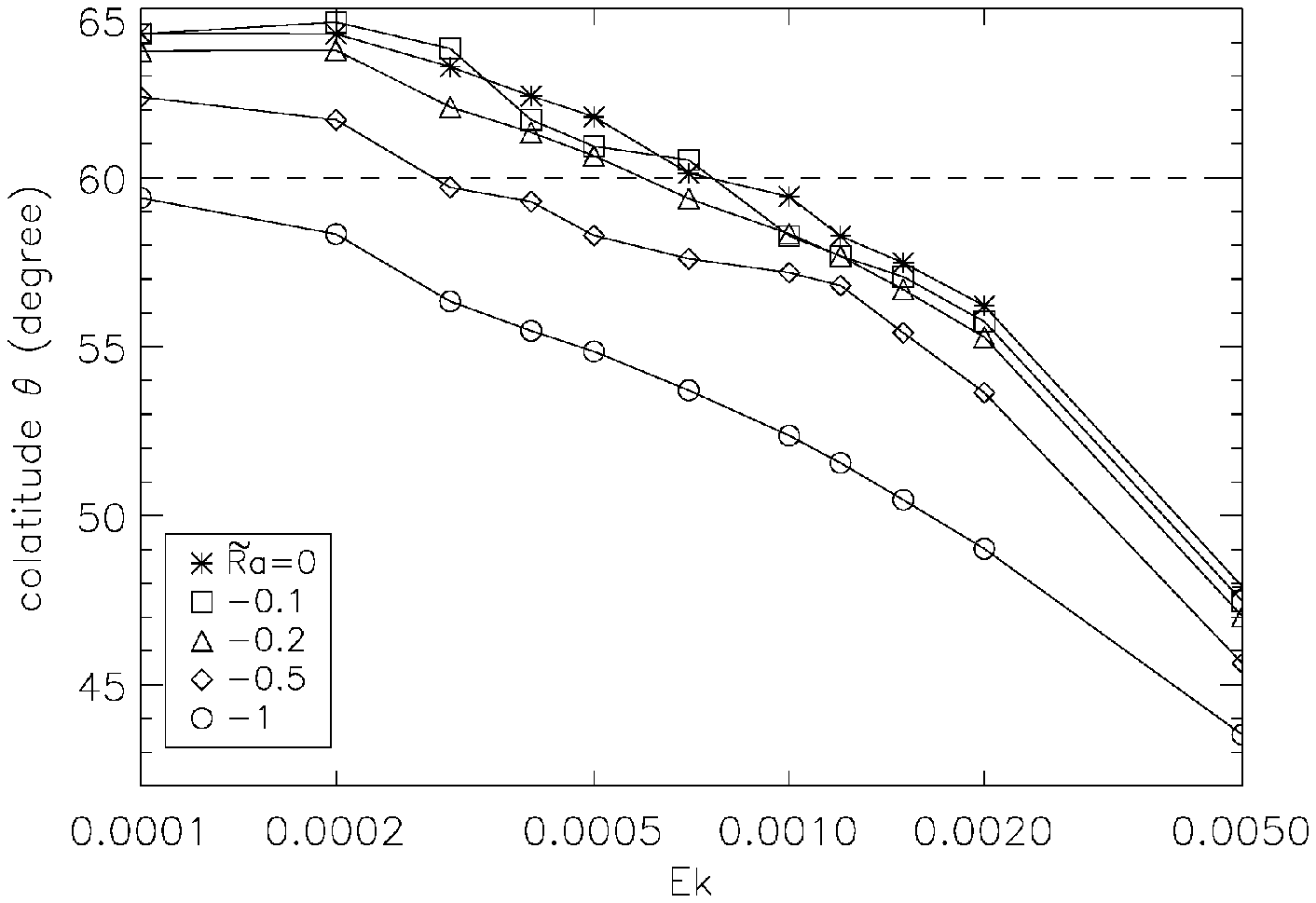}\label{colatitude}}
\subfigure[]
{\includegraphics[scale=0.46]{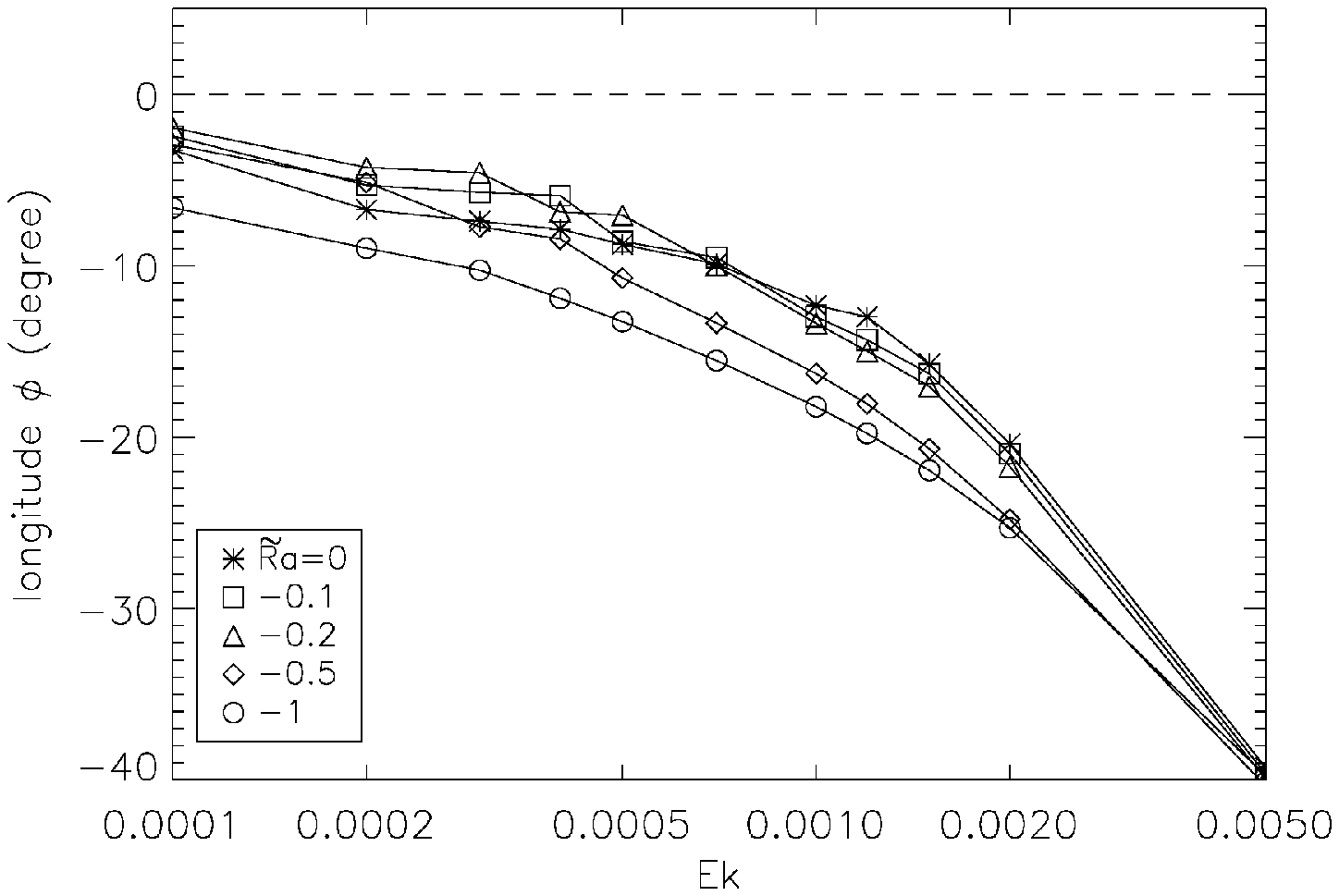}\label{longitude}}
\end{center}
\caption{\footnotesize Stably stratified precessional flow at Poincar\'e number $Po=-0.3$. Position of fluid rotation axis as a function of Ekman number $Ek$ at different Rayleigh numbers $\widetilde{Ra}$. Colatitude (a) and longitude (b) of fluid rotation axis in the precession frame. The dashed line denotes the position of precession axis.}\label{angle}
\end{figure}

The orientation of the fluid rotation axis is best characterised in the precession frame in which the geographic and precession axes are stationary. Figure \ref{angle} shows the angle formed by the rotation axis of the fluid (averaged over the fluid volume) with the geographic axis (figure \ref{colatitude}) and with the meridian of the precession axis (figure \ref{longitude}) in the precession frame. Figure \ref{colatitude} parallels the total kinetic energy shown in figure \ref{E}, since the kinetic energy increases with the increasing angle between the fluid and boundary rotation axes. The key point in figure \ref{colatitude} is that the stable stratification does not noticeably modify the rotation axis until $\widetilde{Ra}=-1$, and therefore we may conclude that a stable stratification takes effect on the precessional flow only for $\widetilde{|Ra|}>1$.

\begin{figure}
\begin{center}
\subfigure[]
{\includegraphics[scale=0.46]{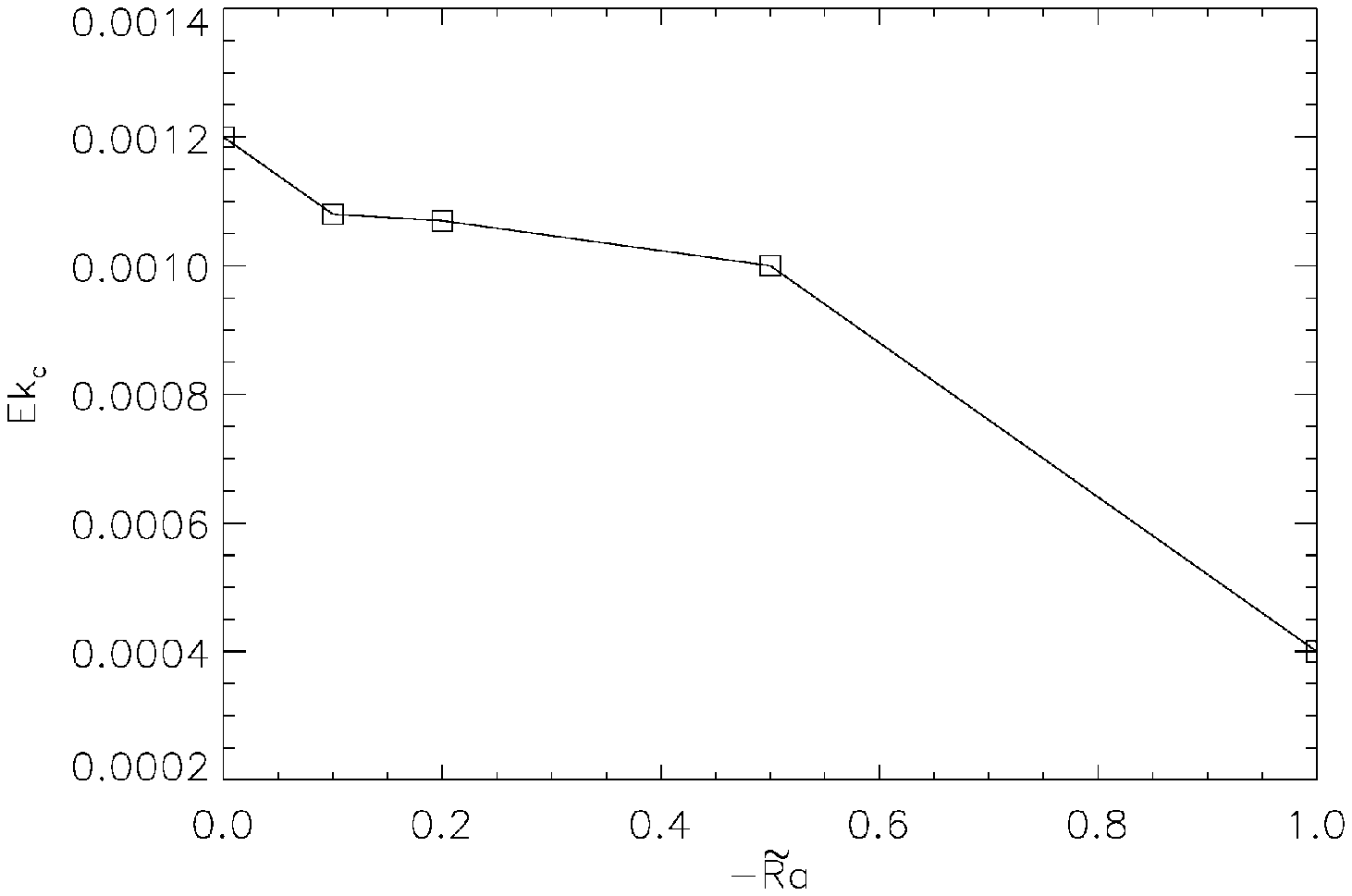}\label{Ek-Ra}}
\subfigure[]
{\includegraphics[scale=0.46]{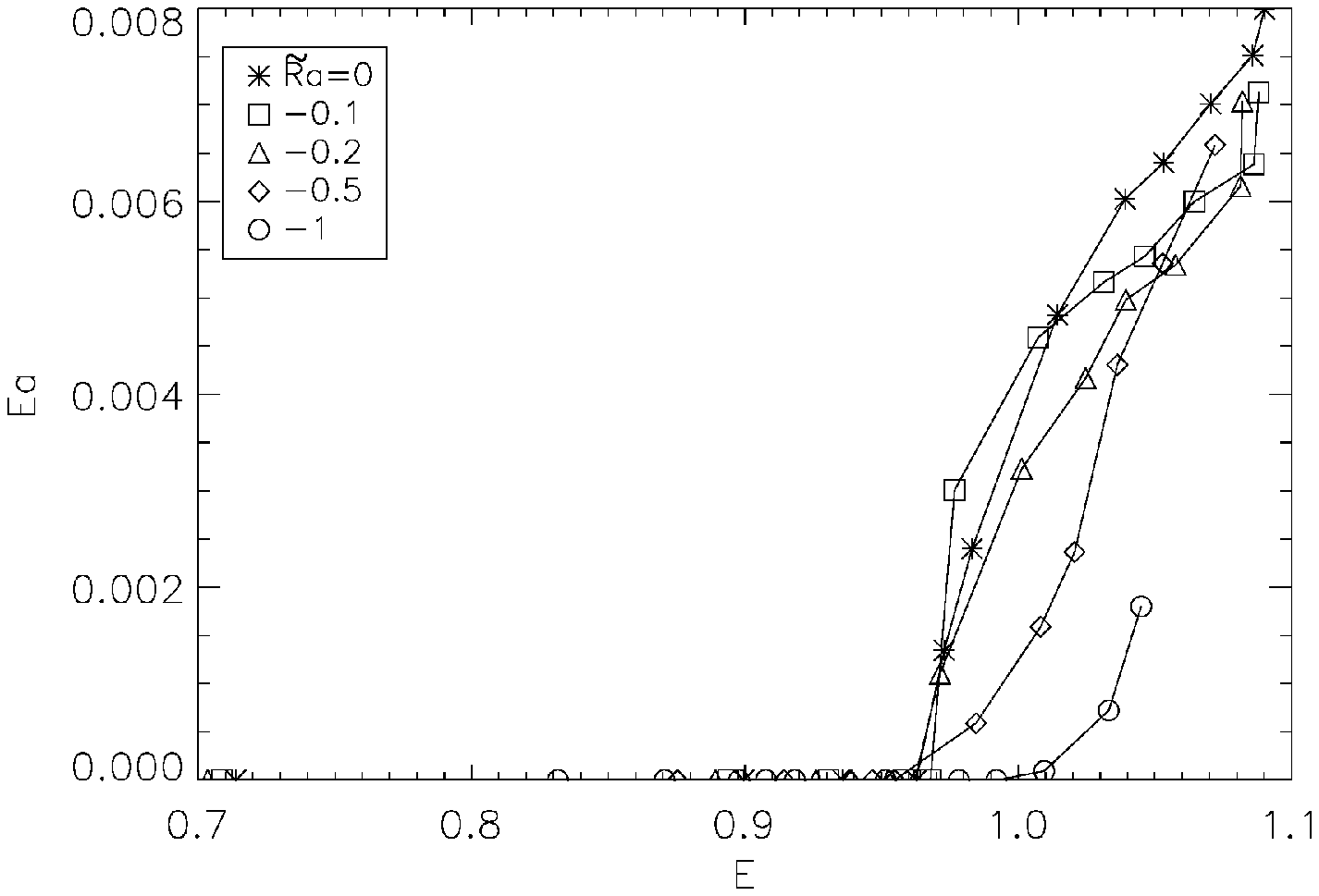}\label{Ea-E}}
\end{center}
\caption{\footnotesize Stably stratified precessional flow at Poincar\'e number $Po=-0.3$. (a) The squares show points determined numerically to lie on the stability limit. The points are connected by a line to guide the eye. (b) The instability energy $Ea$ as a function of the total energy $E$ for different Rayleigh numbers $\widetilde{Ra}$.}
\end{figure}

The energy in the unstable modes is shown in figure \ref{Ea}. $E_a \neq 0$ indicates instability. The onset of instability is shifted towards smaller $Ek$ if $\widetilde{|Ra|}$ is increased. This can be seen in more detail in figure \ref{Ek-Ra} which shows the critical Ekman number, below which the flow is unstable, as a function of $\widetilde{Ra}$. The critical Ekman number monotonously decreases with increasing $\widetilde{|Ra|}$. Even though this is intuitive (because stable stratification suppresses radial motion and hence retards the onset of instability), the opposite could have happened as well if the changes in frequency of the inertial modes due to stratification had brought a combination of them closer to a triad resonance. There is an indication of this effect at finite amplitude, since the hierarchy among the different $\widetilde{Ra}$ is not preserved in figure \ref{Ea} at $Ek=10^{-3}$.

The onset of the instability is to a large degree just a matter of the energy in the basic flow. The growth rate of a triad resonance depends on the shear in the basic flow which increases with the energy of the basic flow. This point is demonstrated in figure \ref{Ea-E} which shows that the onsets of instability for different $\widetilde{Ra}$ nearly coincide if $E_a$ is plotted as a function of $E$, with the exception of $\widetilde{Ra}=-1$. This again shows that stable stratification is significant only for $\widetilde{|Ra|}=1$ or larger.

\section{Unstably stratified precessional flow}
For the study of the fluid core of most planets it is likely more relevant to investigate the interaction of unstable stratification with precession. In \citet{bars2} the geometry of an infinitely long cylinder with an elliptical cross section was analytically investigated and it was shown that the elliptical instability and the convective instability may either stabilise or destabilise each other in different parameter regimes. In this section we numerically calculate unstably stratified precessional flow in spherical geometry to study the interaction of precession with unstable stratification.

We select two  Ekman numbers, $Ek=5\times10^{-3}$ and $1\times10^{-3}$, such that the purely precessional flow is stable, namely it is stable at least until $|Po|=1$ for $Ek=5\times10^{-3}$ and $|Po|=0.3$ for $Ek=1\times10^{-3}$. Then we test various combinations of $Po$ and $\widetilde{Ra}$ to seek the neutral stablity curves of the critical Rayleigh number $\widetilde{Ra}_c$ against $|Po|$ at both Ekman numbers, as well as to calculate some supercritical flows at $Ek=5\times10^{-3}$. $\widetilde{Ra}_c$ is sought by increasing or decreasing $\widetilde{Ra}$ in steps of $0.01$ at a given $Po$ . We distinguish two different types of flows by the dominant azimuthal wavenumbers in the spectra of the velocity field. These spectra are computed in a frame of reference in which the axes of precession and boundary rotation are stationary and a $z'$-axis is pointing along the rotation axis of the fluid. The azimuthal angle $\phi'$ is measured around the $z'$-axis and the azimuthal wavenumbers are denoted by $m'$. The onset of convection occurs at $m'=3$, so that we call the flow convective if the kinetic energy is concentrated at $m'=3$. If on the contrary the spectrum has its largest contributions at $m'<3$, we call the flow precessional. Flows undergoing precessional instability have their kinetic energy concentrated in these wavenumbers. The motion with wavenumbers $m'=1$ and $2$ can be identified as two inertial modes by the same technique as used by \citet{tilgner7}. These two modes, together with the spin-over mode, fulfill the requirements of a triad resonance. Their parameters (latitudinal wavenumber $l'$ and frequency $\varpi'$) are close to two analytically determined eigenmodes with $m'=1$, $l'=6$, $\varpi'=-0.537$ and $m'=2$, $l'=6$, $\varpi'=-1.093$ \citep{greenspan}.

\begin{figure}
\begin{center}
\subfigure[]
{\includegraphics[scale=0.48]{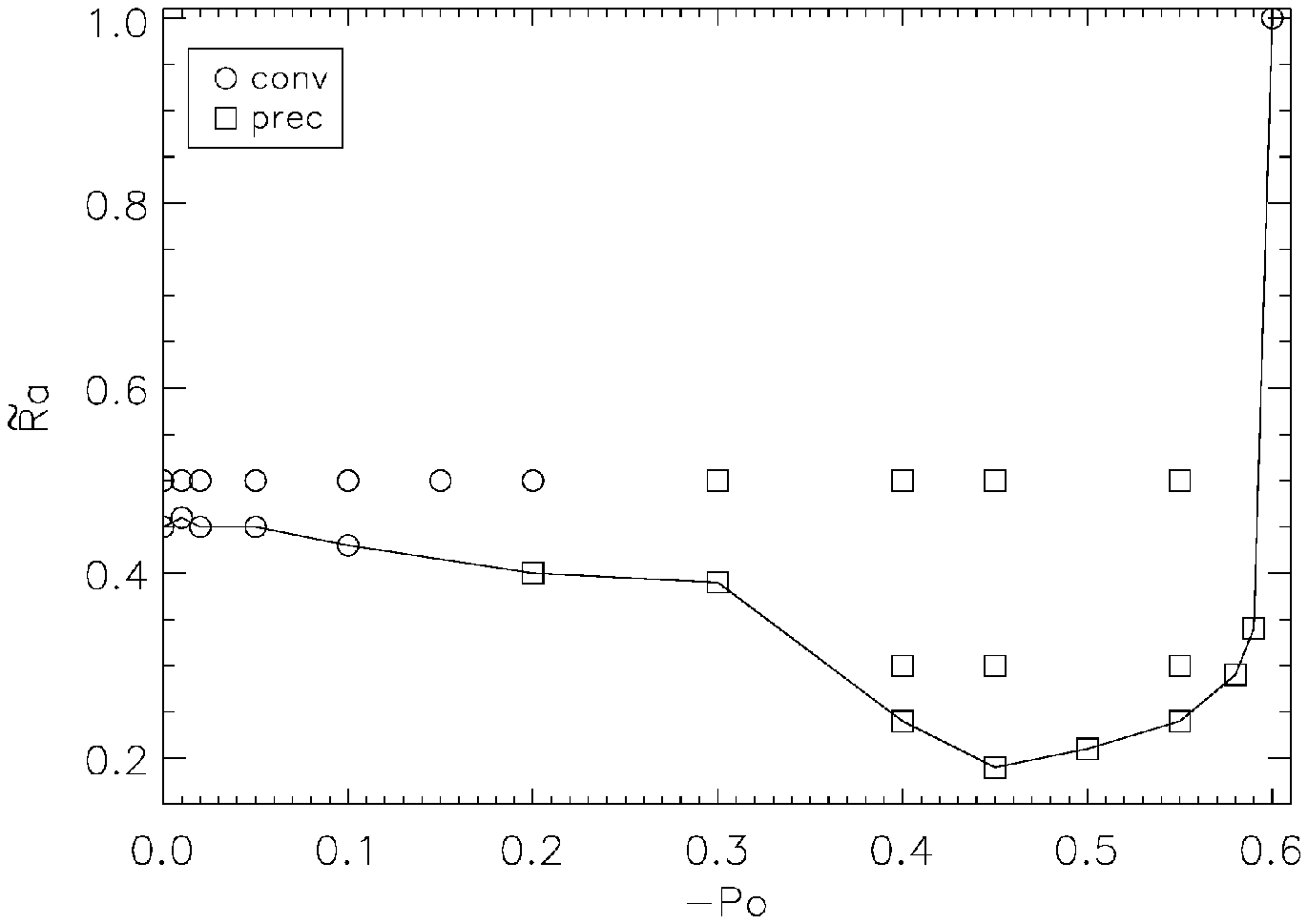}\label{RaPo1}}
\subfigure[]
{\includegraphics[scale=0.48]{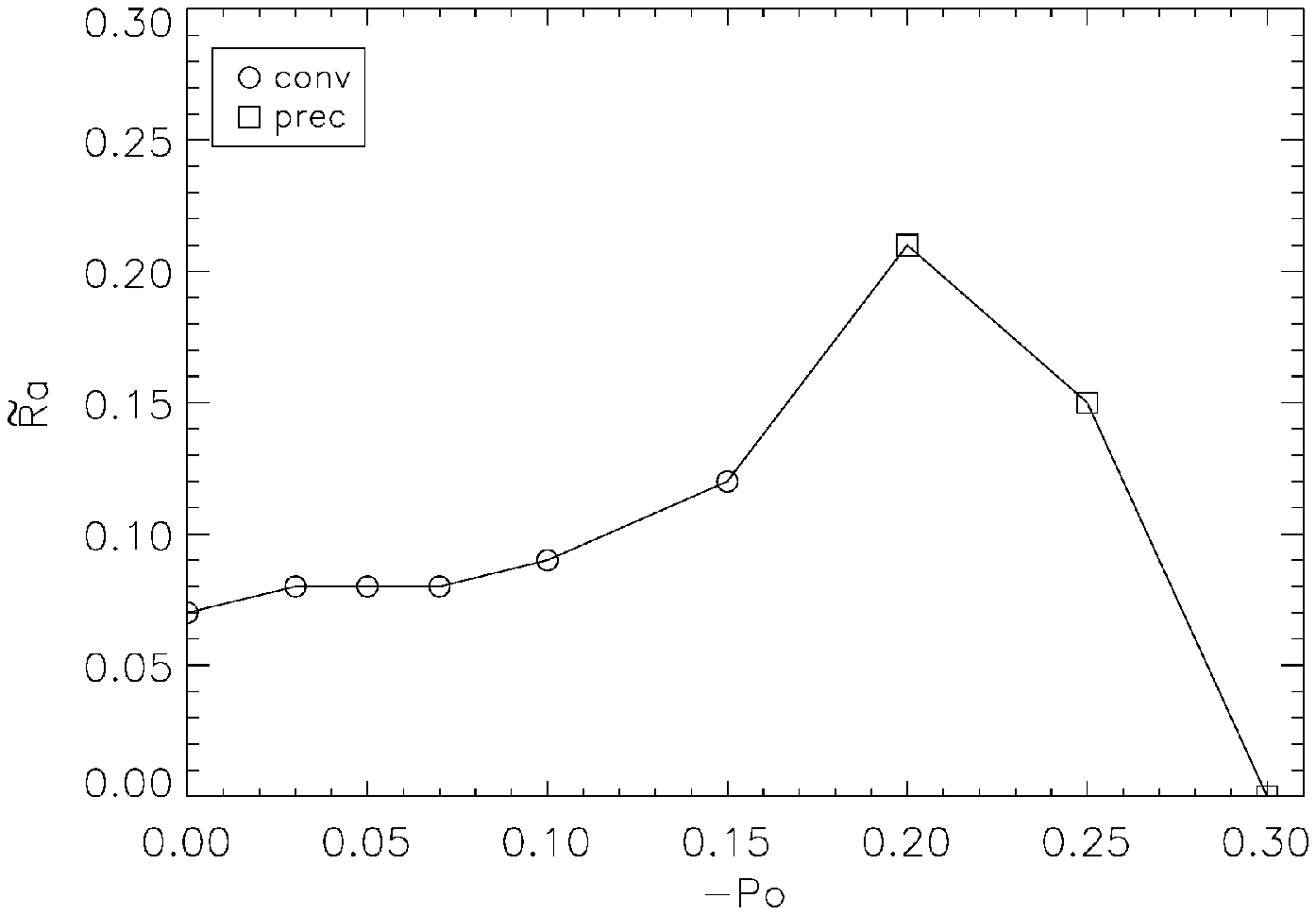}\label{RaPo2}}
\end{center}
\caption{\footnotesize Diagram of convective stability and precessional stability for unstably stratified flow at $Ek=5\times10^{-3}$ (a) and $Ek=1\times10^{-3}$ (b). Points on the solid line denotes points on the neutral stability curve. The circle symbols denote the convective flows and the square symbols the precessional flows.}\label{RaPo}
\end{figure}

Figure \ref{RaPo1} shows an approximative stability diagram of unstably stratified precessional flows at $Ek=5\times10^{-3}$. The solid line denotes the neutral stability curve above/below which flow is unstable/stable, and the circle symbol denotes the convective flows and the square symbol the precessional flows. In the absence of precession ($Po=0$), $\widetilde{Ra}_c$ for the onset of convective instability is $0.45$ and the dominant mode is $m'=3$. When $|Po|$ increases to $0.1$, $\widetilde{Ra}_c$ decreases and the instability is still convective. The instability becomes precessional for $|Po|=0.2$ or larger. $\widetilde{Ra}_c$ decreases until it reaches a minimum of $\widetilde{Ra}_c=0.19$ at  $|Po|=0.45$. The flow becomes more stable with stronger precession for $|Po|>0.45$. At $|Po|=0.6$ the flow is so stable that $\widetilde{Ra}_c$ reaches $1$. At such a high $\widetilde{Ra}$ the flow pattern is convective again.

Precession has apparently a dual role. On one hand, both precession and convection can lead to instabilities on their own so that a superposition of both can be expected to be even less stable. This is observed in figure \ref{RaPo1} around $-Po \approx 0.45$, where the critical Rayleigh number is reduced by more than a factor of 2 by precession, but precession alone in the isothermal fluid (corresponding to $\widetilde{Ra}=0$) is stable. On the other hand, precession introduces shear on top of the global rotation, and the onset of convective instability has now to be computed for a sheared basic flow, which can lead to a higher critical Rayleigh number. An example of this phenomenon is seen in figure \ref{RaPo1} for $-Po$ around 0.6.

We then investigate the lower Ekman number of  $Ek=1\times10^{-3}$. We only seek the neutral stability curve but do not calculate any supercritical flows, and we calculate until $|Po|=0.3$ beyond which the purely precessional flow ($\widetilde{Ra}=0$) is already unstable (figure \ref{Ea}). Figure \ref{RaPo2} shows this neutral stability curve. As in figure \ref{RaPo1} the circle symbol denotes the convective instability and the square symbol the precessional instability. For purely convective flows ($Po=0$), the critical Rayleigh number $\widetilde{Ra}_c=0.07$ at $Ek=1\times10^{-3}$ is much lower than $\widetilde{Ra}_c=0.45$ at $Ek=5\times10^{-3}$. This is consistent with the asymptotic scaling law of previous calculations \citep{roberts2, busse3, zhang} and numerical calculations \citep{zhang, tilgner6}. Notice that the definition of $Ra$ in \citep{tilgner6} should be translated to our definition (equation \ref{ra}) such that the asymptotic scaling law is $\widetilde{Ra}_c=O(Ek^{2/3})$ (this scaling law does not precisely hold in our numerical calculations because we choose too high Ekman numbers). For $Ek=1\times10^{-3}$, $\widetilde{Ra}_c$ has a maximum value of $\widetilde{Ra}_c=0.21$ at $|Po|=0.2$, which shows that precession stabilises the flow at small $|Po|$ whereas it destabilises at large $|Po|$. We know that the neutral stability curve at $Ek=5\times10^{-3}$ (figure \ref{RaPo1}) eventually decreases when $|Po|$ is large enough for precession alone to be unstable. In summary, at both the high and low Ekman numbers the neutral stability curve is not monotonic but it has a minimum at the high $Ek$ and a maximum at the low $Ek$.

\begin{figure}
\begin{center}
\includegraphics[scale=0.48]{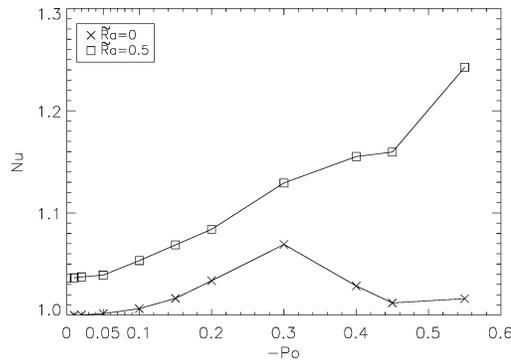}
\end{center}
\caption{\footnotesize Nusselt number $Nu$ at the outer boundary as a function of $Po$ at Ekman number $Ek=5\times10^{-3}$ and Rayleigh numbers $\widetilde{Ra}=0$ and $0.5$.}\label{nu}
\end{figure}

To end this section we discuss the Nusselt number $Nu$ which measures the ratio of the total heat transfer to the thermal conduction in the fluid at rest. The stratifying linear temperature profile assumed here is maintained by a heat source so that the Nusselt number depends on radius.
The Nusselt number is conveniently computed at the boundaries as
\begin{equation}
Nu=\frac{\langle \partial T/\partial r \rangle}{\Delta T/d}.
\end{equation}
The brackets denote an average over the spherical surface of radius $r_o$ or $r_i$ and time. Because of the internal heat source, the Nusselt numbers at the two boundaries are related through
\begin{equation}\label{twonu}
Nu|_{r=r_o}-\eta^2Nu|_{r=r_i}=1-\eta^2,
\end{equation}
where $\eta=r_i/r_o=0.1$ is the aspect ratio used in our calculations.
It is verified by our numerical calculations that equation (\ref{twonu}) precisely holds. Since $Nu$ at the inner boundary can be deduced from the one at the outer boundary, we only show $Nu$ at the outer boundary. Figure \ref{nu} shows $Nu$ at the outer boundary against $Po$ at $Ek=5\times10^{-3}$ for $\widetilde{Ra}=0$ and $\widetilde{Ra}=0.5$. Because a purely precessional flow ($\widetilde{Ra}=0$) has some poloidal component it also transfers heat, in which case the temperature deviation is a passive scalar. Figure \ref{nu} contains several points with $Nu$ around $1.05$ which are compared in table \ref{urtheta}. At equal poloidal kinetic energy, one flow may be more efficient at tranporting heat than the other because of a better correlation between radial velocity and temperature. According to this criterion, convection is more efficient than precession at advecting heat. Adding precession to a convective flow reduces this correlation by a factor of about $4$, as shown in table \ref{urtheta}.

\begin{table}
\begin{center}
\begin{tabular*}{\textwidth}{@{\extracolsep{\fill}}lllllll}
$\widetilde{Ra}$ & $Po$ & $Nu$ & $\langle\int u_r\Theta dV\rangle$ & $\langle\int u_r^2dV\rangle$ & $\langle\int\Theta^2dV\rangle$ & $\frac{\langle\int u_r\Theta dV\rangle}{\sqrt{\langle\int u_r^2dV\rangle\langle\int\Theta^2dV\rangle}}$ \\
$0$   & $-0.1$ & $1.006$ & $6.08\times10^{-4}$ & $3.81\times10^{-3}$ & $4.22\times10^{-3}$ & $0.151$ \\
$0$   & $-0.2$ & $1.034$ & $2.57\times10^{-3}$ & $1.33\times10^{-2}$ & $1.91\times10^{-2}$ & $0.161$ \\
$0$   & $-0.3$ & $1.069$ & $4.24\times10^{-3}$ & $1.50\times10^{-2}$ & $2.76\times10^{-2}$ & $0.209$ \\
$0$   & $-0.4$ & $1.029$ & $2.07\times10^{-3}$ & $4.78\times10^{-3}$ & $1.10\times10^{-2}$ & $0.285$ \\
$0.5$ & $0$    & $1.036$ & $2.04\times10^{-3}$ & $4.73\times10^{-4}$ & $1.29\times10^{-2}$ & $0.826$ \\
$0.5$ & $-0.1$ & $1.053$ & $3.07\times10^{-3}$ & $4.60\times10^{-3}$ & $2.07\times10^{-2}$ & $0.315$ \\
$0.5$ & $-0.2$ & $1.084$ & $4.81\times10^{-3}$ & $1.56\times10^{-2}$ & $3.40\times10^{-2}$ & $0.209$
\end{tabular*}
\end{center}
\caption{\footnotesize Correlations $\langle\int u_r\Theta dV\rangle$, $\langle\int u_r^2dV\rangle$ and $\langle\int\Theta^2dV\rangle$ of purely precessional and convective flows. The brackets denote the time average. The first four flows at $\widetilde{Ra}=0$ are precessionally stable and the next three flows at $\widetilde{Ra}=0.5$ are convectively unstable as shown in figure \ref{RaPo1}.}\label{urtheta}
\end{table}

\section{Discussion}
We investigated the interaction of precession with thermal stratification. The heat transport in precessional flow is less than in convective flow at equal rms of the radial velocity component due to a smaller correlation between radial velocity and temperature. Unstable stratification together with precession can be either more stable or more unstable than the stratification or precession acting alone. The same could have been expected from the combination of precession with stable stratification, but in this case, we only found examples in which the stratification stabilises the flow. The presence of stable stratification becomes relevant for precession if the ratio of the Brunt-V\"ais\"al\"a frequency to the rotation frequency is near $1$.

We present a rough estimate to determine what this criterion implies for various celestial bodies. Consider the extreme case of a body which has cooled down so much that it is nearly isothermal. Within the Boussinesq approximation used here, this corresponds to a stable stratification with a temperature gradient equal to the adiabatic gradient, given by
\begin{equation}
\left(\frac{\partial T}{\partial z}\right)_{\rm ad}=-\frac{g\alpha T}{c_p},
\end{equation}
where $g$ is the gravitational acceleration, $\alpha$ the thermal expansion coefficient and $c_p$ the specific heat capacity at constant pressure.
The gravitational accleration at the surface of a sphere with radius $R$ is given by
\begin{equation}
g=\frac{4}{3}\pi G\rho R,
\end{equation}
where $G$ is the gravitational constant. The Brunt-V\"ais\"al\"a frequency for the cooled body under consideration is given by
\begin{equation}\label{N}
N=\sqrt{-g\alpha\left(\frac{\partial T}{\partial z}\right)_{\rm ad}}=g\alpha\sqrt{\frac{T}{c_p}}=\frac{4}{3}\pi G\rho\alpha R\sqrt{\frac{T}{c_p}}.
\end{equation}
If we accept the following values to be representative of planetary cores
\begin{equation}\label{values}
\rho=11390\,{\rm kg/m^3},\hspace{3mm}\alpha=1.5\times10^{-5}\,{\rm K^{-1}},\hspace{3mm}T=4000\,{\rm K},\hspace{3mm}c_p=860\,{\rm J/(kg\cdot K)},
\end{equation}
we obtain
\begin{equation}
N=1.03\times10^{-10}R\;{\rm m^{-1}s^{-1}}.
\end{equation}
If we take as an example planetesimals with a  typical radius of $R=10\,{\rm km}$  or $R=100\,{\rm km}$, we find  $N=10^{-6}\,{\rm s^{-1}}$ or  $N=10^{-5}\,{\rm s^{-1}}$, respectively. The rotation of planetesimals has a typical period of about $10$ hours \citep{weiss}, i.e. the rotation rate is $\Omega=1.7\times10^{-4}\,{\rm s^{-1}}$, which implies that the buoyancy is negligible in this context. The situation is different for the Earth with a core of radius $3.4 \times 10^6 {\rm m}$ and a rotation period of $24$ hours. A stabilising stratification with a gradient $4\%$ less than the adiabatic gradient would be enough for the stratification to significantly influence the flow driven by precession.

%{\small\section*{Acknowledgement}This work is financially supported by the project SPP1488 of the program PlanetMag of Deutsche Forschungsgemeinschaft (DFG). We thank the Leibniz-Institute Astrophysics Potsdam to host Xing Wei as a visiting researcher.}

\bibliographystyle{jfm}
\bibliography{paper}

\begin{thebibliography}{27}
\expandafter\ifx\csname natexlab\endcsname\relax\def\natexlab#1{#1}\fi

\bibitem[Bullard(1949)]{bullard}
{\sc Bullard, E.~C.} 1949 The magnetic field within the earth. {\em Proc. Roy.
  Soc. A\/} {\bf 197}, 433--453.

\bibitem[Busse(1968)]{busse2}
{\sc Busse, F.~H.} 1968 Steady fluid flow in a precessing spheroidal shell.
  {\em J. Fluid Mech.\/} {\bf 33}, 739--751.

\bibitem[Busse(1970)]{busse3}
{\sc Busse, F.~H.} 1970 Thermal instabilities in rapidly rotating systems. {\em
  J. Fluid Mech.\/} {\bf 44}, 441--460.

\bibitem[Busse(2000)]{busse1}
{\sc Busse, F.~H.} 2000 Homogeneous dynamos in the planetary cores and in the
  laboratory. {\em Annu. Rev. of Fluid Mech.\/} {\bf 32}, 383--408.

\bibitem[Dwyer {\em et~al.\/}(2011)Dwyer, Stevenson \& Nimmo]{dwyer}
{\sc Dwyer, C.~A., Stevenson, D.~J. \& Nimmo, F.} 2011 A long-lived lunar
  dynamo driven by continuous mechanical stirring. {\em Nature\/} {\bf 479},
  212--215.

\bibitem[Greenspan(1968)]{greenspan}
{\sc Greenspan, H.~P.} 1968 {\em The theory of rotating fluids\/}, 1st edn.
  Cambridge, U.K.: Cambridge University Press.

\bibitem[Jones(2011)]{jones}
{\sc Jones, C.~A.} 2011 Planetary magnetic fields and fluid dynamos. {\em Annu.
  Rev. of Fluid Mech.\/} {\bf 43}, 583--614.

\bibitem[Kerswell(1993)]{kerswell2}
{\sc Kerswell, R.} 1993 Elliptical instabilities of stratified hydromagnetic
  waves. {\em Geophys. Astrophys. Fluid Dyn.\/} {\bf 71}, 105--143.

\bibitem[Kerswell(1995)]{kerswell1}
{\sc Kerswell, R.} 1995 On the internal shear layers spawned by the critical
  regions in oscillatory ekman boundary layers. {\em J. Fluid Mech.\/} {\bf
  298}, 311--325.

\bibitem[{Le Bars} \& {Le Diz\'es}(2006)]{bars2}
{\sc {Le Bars}, M. \& {Le Diz\'es}, S.} 2006 Thermo-elliptical instability in a
  rotating cylindrical shell. {\em J. Fluid Mech.\/} {\bf 563}, 189--198.

\bibitem[{Le Bars} {\em et~al.\/}(2011){Le Bars}, Wieczorek, Karatekin,
  C\'ebron \& Laneuville]{bars1}
{\sc {Le Bars}, M., Wieczorek, M.~A., Karatekin, O., C\'ebron, D. \&
  Laneuville, M.} 2011 An impact-driven dynamo for the early moon. {\em
  Nature\/} {\bf 479}, 215--218.

\bibitem[Lorenzani \& Tilgner(2001)]{tilgner7}
{\sc Lorenzani, S. \& Tilgner, A.} 2001 Fluid instabilitis in precessing
  spheroidal cavities. {\em J. Fluid Mech.\/} {\bf 447}, 111--128.

\bibitem[Lorenzani \& Tilgner(2003)]{tilgner3}
{\sc Lorenzani, S. \& Tilgner, A.} 2003 Inertial instabilities of fluid flow in
  precessing spheroidal shells. {\em J. Fluid Mech.\/} {\bf 492}, 363--379.

\bibitem[Malkus(1968)]{malkus}
{\sc Malkus, V.~R.} 1968 Precession of the earth as the cause of geomagnetism.
  {\em Science\/} {\bf 160}, 259--264.

\bibitem[Noir {\em et~al.\/}(2001)Noir, Brito, Aldridge \& Cardin]{noir}
{\sc Noir, J., Brito, D., Aldridge, K. \& Cardin, P.} 2001 Experimental
  evidence of inertial waves in a precessing spheroidal cavity. {\em Geophys.
  R. Lett.\/} {\bf 28}, 3785--3788.

\bibitem[Poincar\'e(1910)]{poincare}
{\sc Poincar\'e, H.} 1910 Sur la pr\'ecession des corps d\'eformables. {\em
  Bulletin Astronomieque\/} {\bf 27}, 321--356.

\bibitem[Roberts(1968)]{roberts2}
{\sc Roberts, P.~H.} 1968 On the thermal instability of a rotating-fluid sphere
  containing heat sources. {\em Phil. Trans. R. Soc. Lond. A\/} {\bf 263},
  93--117.

\bibitem[Roberts \& Glatzmaier(2000)]{roberts1}
{\sc Roberts, P.~H. \& Glatzmaier, G.~A.} 2000 Geodynamo theory and
  simulations. {\em Rev. Mod. Phys.\/} {\bf 72}, 1081--1123.

\bibitem[Tilgner(1999{\natexlab{{\em a\/}}})]{tilgner4}
{\sc Tilgner, A.} 1999{\natexlab{{\em a\/}}} Magnetohydrodynamic flow in
  precessing spherical shells. {\em J. Fluid Mech.\/} {\bf 379}, 303--318.

\bibitem[Tilgner(1999{\natexlab{{\em b\/}}})]{tilgner5}
{\sc Tilgner, A.} 1999{\natexlab{{\em b\/}}} Spectral methods for the
  simulation of incompressible flows in spherical shells. {\em Int. J. Numer.
  Meth. Fluids\/} {\bf 30}, 713--724.

\bibitem[Tilgner(2005)]{tilgner2}
{\sc Tilgner, A.} 2005 Precession driven dynamo. {\em Phys. Fluids\/} {\bf 17},
  034104.

\bibitem[Tilgner \& Busse(1997)]{tilgner6}
{\sc Tilgner, A. \& Busse, F.~H.} 1997 Finite-amplitude convection in rotating
  spherical fluid shells. {\em J. Fluid Mech.\/} {\bf 332}, 359--376.

\bibitem[Tilgner \& Busse(2001)]{tilgner1}
{\sc Tilgner, A. \& Busse, F.~H.} 2001 Fluid flows in precessing spherical
  shells. {\em J. Fluid Mech.\/} {\bf 426}, 387--396.

\bibitem[Vanyo \& Likins(1971)]{vanyo}
{\sc Vanyo, J. \& Likins, P.} 1971 Measurement of energy dissipation in a
  liquid-filled, precessing, spherical cavity. {\em ASME Trans. J. Appl.
  Mech.\/} {\bf 38}, 674--682.

\bibitem[Weiss {\em et~al.\/}(2008)Weiss, Berdahl, Tanton, Stanley, Lima \&
  Carporzen]{weiss}
{\sc Weiss, B.~P., Berdahl, J.~S., Tanton, L.~E., Stanley, S., Lima, E.~A. \&
  Carporzen, L.} 2008 Magnetism on the angrite parent body and the early
  differentiation of planetesimals. {\em Science\/} {\bf 322}, 713--716.

\bibitem[Wicht \& Tilgner(2010)]{wicht}
{\sc Wicht, J. \& Tilgner, A.} 2010 Theory and modeling of planetary dynamos.
  {\em Space Sci. Rev.\/} {\bf 152}, 501--542.

\bibitem[Zhang {\em et~al.\/}(2007)Zhang, Liao \& Busse]{zhang}
{\sc Zhang, K., Liao, X. \& Busse, F.~H.} 2007 Asymptotic solutions of
  convection in rapidly rotating no-slip spheres. {\em J. Fluid Mech.\/} {\bf
  578}, 371--380.

\end{thebibliography}

\end{document}